\documentclass[aps,pra,showpacs,11pt]{revtex4-2}
\usepackage[english]{babel}
\usepackage[utf8]{inputenc}
\usepackage{graphicx}
\usepackage{amssymb}
\usepackage{xcolor}
\usepackage{amsmath}
\usepackage[utf8]{inputenc}
\usepackage{hyperref}
\usepackage{url}
\usepackage{color}
\usepackage{enumerate}

\begin{document}
%
\title{Optimización de la Transmisión de Estados Cuánticos en Cadenas de Qubits usando Deep Reinforcement Learning y Algoritmos Genéticos.}

\author{
	Sofía Perón Santana  (0009-0005-1545-0690) $^{1,2, *}$,
	Ariel Fiuri (0009-0008-9762-9134) $^{2}$,
	Omar Osenda (0000-0003-3251-7031) $^{1,2}$, 
	Martín Domínguez  (0000-0002-1815-0117) $^{2}$\\[6pt]
	\small
	$^{1}$Instituto de Física Enrique Gaviola (CONICET-UNC), Córdoba, Argentina\\
	$^{2}$Facultad de Matemática, Astronomía, Física y Computación (FAMAF), Universidad Nacional de Córdoba, Córdoba, Argentina\\
	$^{*}$Contacto: sofia.peron@mi.unc.edu.ar
}


\begin{abstract}
	La transferencia de estado cuántico (QST) a través de cadenas de espín homogéneas desempeña un papel crucial en la construcción de hardware cuántico escalable. Un protocolo básico de transmisión de estados cuánticos prepara un estado en un qubit y lo transfiere a otro a través de un canal, buscando minimizar el tiempo y evitar pérdida de información. La fidelidad del proceso se mide mediante funciones proporcionales a la probabilidad de transición entre ambos estados. Abordamos este problema de optimización mediante pulsos magnéticos constantes y dos estrategias complementarias: aprendizaje por refuerzo profundo, donde un agente aprende secuencias de pulsos mediante recompensas, y algoritmos genéticos, que desarrollan soluciones candidatas mediante selección y mutación. Analizamos la eficiencia de ambos métodos y su capacidad para incorporar restricciones físicas.

	Quantum state transfer (QST) via homogeneous spin chains plays a crucial role in building scalable quantum hardware. A basic quantum state transmission protocol prepares a state in one qubit and transfers it to another through a channel, seeking to minimize the time and avoid information loss. The fidelity of the process is measured by functions proportional to the transition probability between both states. We approach this optimization problem using constant magnetic pulses and two complementary strategies: deep reinforcement learning, where an agent learns pulse sequences through rewards, and genetic algorithms, which develop candidate solutions through selection and mutation. We analyze the efficiency of both methods and their ability to incorporate physical constraints.
\end{abstract}

\maketitle

\section{Introducción}

La transmisión de estados cuánticos es un área que lleva más de veinte años de desarrollo (\cite{Peron2025} y las referencias citadas). El protocolo básico es muy simple, se prepara un vector de estado inicial en un extremo de una cadena de qubits y se pretende obtener dicho estado en el otro extremo de la cadena. En Mecánica Cuántica, por vector de estado se entiende un vector {\em complejo} que pertenece a un espacio vectorial, el cual contiene todos los estados posibles. El estado inicial evoluciona siguiendo el {\bf Hamiltoniano} del sistema, el cual contiene todas las interacciones entre los elementos de la cadena y los controles externos aplicados a la misma. Las interacciones dependen de la física del problema, pero en general se pueden escribir como las interacciones entre dipolos magnéticos. 

La ecuación que gobierna la evolución temporal de un vector de estado $\psi(t)$ es la ecuación de Schrödinger
\begin{equation}
i \frac{d \psi(t)}{d t} = H \psi(t),
\end{equation}
donde el Hamiltoniano $H$ es una matriz. Si el Hamiltoniano es independiente del tiempo, la solución es trivial
\begin{equation}
\psi(t) = \exp{(i H t)} \psi(0) = U(t)\psi(0)  . 
\end{equation}
Si el Hamiltoniano depende del tiempo, pero sólo a través de cantidades que son constantes en el tiempo durante un intervalo, la solución se puede generalizar en forma simple
\begin{eqnarray}
\psi(t) &=& \exp{(i H_n \tau_n)} \exp{(i H_{n-1} \tau_{n-1} )}\ldots \exp{(i H_2 \tau_2)}\exp{(i H_1 \tau_1)} \psi(0) \\
        &=& U_n U_{n-1} \ldots U_2 U_1 \psi(0) = U \psi(0)
\end{eqnarray}
donde $H_n$ es el Hamiltoniano independiente del tiempo válido en el intervalo de tiempo $\tau_n$ y $t$ es el tiempo $t=\tau_1 + \tau_2 + \ldots +\tau_n$.

\begin{figure}
    \centering
    \includegraphics[width=0.92\textwidth]{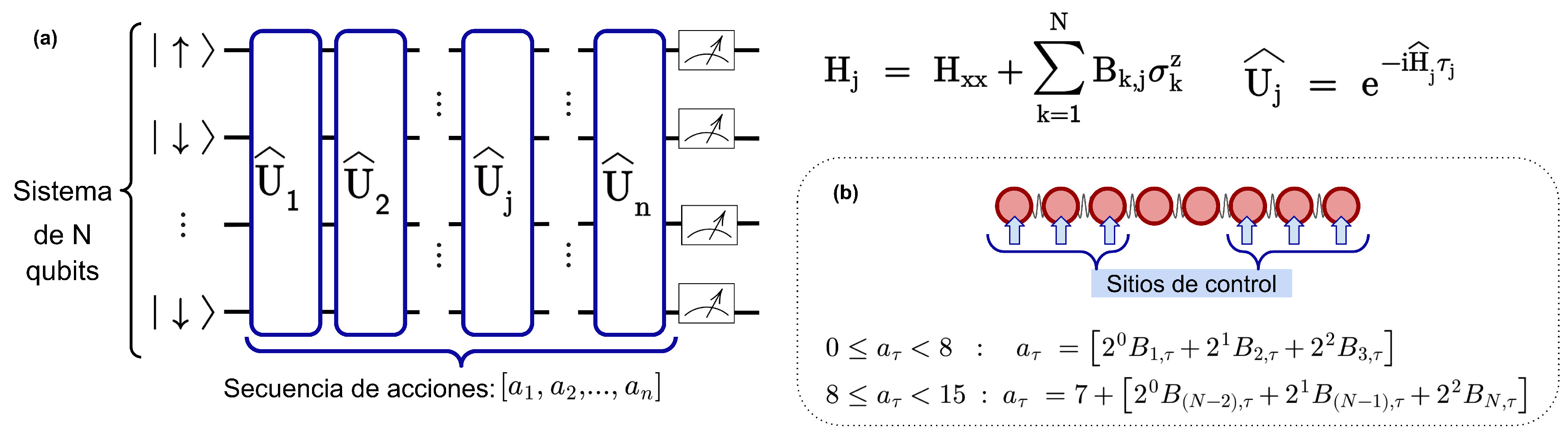}
    \caption{(a) Representación en forma de circuito cuántico del protocolo de QST. Se inicializa un estado de N qubits con una única excitación y se aplica una serie de \textit{acciones} $[a_1,a_2,...,a_A]$ que representan compuertas actuando sobre los distintos qubits. (b) Esquema de los sitios de control con los que se efectúan las acciones sobre el sistema. Si $B_{jk}$ es distinto de cero esto implica que el control sobre el k-ésimo qubit en el j-ésimo intervalo de tiempo está encendido.}
    \label{fig:circuito}
    \end{figure}

Vamos a considerar que el Hamiltoniano se puede escribir como
\begin{equation}
H_j = \sum_{i=1}^{N-1} (\sigma_i^x \sigma_{i+1}^x+\sigma_i^y \sigma_{i+1}^y ) + \sum_{k=1}^N B_{k,j} \sigma_k^z = H_{XX} + \sum_{k=1}^N B_{k,j} \sigma_k^z,
\end{equation}
donde $B_{k,j}$ es el control que se aplica sobre el qubit $k-$ésimo, las matrices complejas $\sigma_i^x, \sigma_i^y, \sigma_i^z$ son las matrices de Pauli. Para convertir el Hamiltoniano en una matriz de dimensión $2^N \times 2^N$ se usa el producto tensorial usual. Los controles $B_{jk}$ pueden configurarse de distinta forma, aquí estudiaremos la implementación de \cite{Zhang2018}, que considera controles en los primeros y últimos tres elementos de la cadena. Esto da un total de 16 acciones posibles: todos los controles distintos de cero, las combinaciones de  1,2 o 3 controles distintos de cero en alguno de los dos extremos y todos los controles iguales a cero. En la figura \ref{fig:circuito}a, se muestra en forma de circuito cuántico el protocolo de transmisión y se denota cada operación $U_k$ como una compuerta que actúa sobre todos los qubits. Un esquema de los sitios de control se muestra en el apartado (b). Note que los campos de control son las {\em acciones}, tanto el algoritmo genético como el aprendizaje reforzado profundo son usados para determinar secuencias de acciones para obtener el vector de estado inicial en el otro extremo de la cadena. 


Si el vector de estado inicial preparado en un extremo de la cadena es $\psi_0$ y el vector de estado {\em target} al final de las acciones está dado por  $\psi_f$, transmitir con alta fidelidad significa que hay que maximizar la probabilidad
\begin{equation}\label{trans_prob}
P = |(\psi_f ,U \psi_0)|^2 ,
\end{equation}
donde se usa el producto interno usual entre matrices y vectores complejos. 

\section{Métodos}

\noindent{\bf Aprendizaje por refuerzo}
\vspace{4pt}

En los algoritmos de aprendizaje por refuerzo (RL), un \textit{agente} aprende por medio de \textit{recompensas} que premian o castigan la capacidad para realizar determinada tarea (\cite{drl}). 

En nuestro trabajo implementamos un modelo del problema físico, el que a su vez constituye el \textit{entorno} con el cual el agente interactúa para generar experiencia y aprendizaje a partir de esta. Este agente tiene la posibilidad de aplicar \textit{acciones} (descriptas en la sección anterior) al entorno y obtiene en contrapartida información de \textit{estado} del mismo y la \textit{recompensa} obtenida por la aplicación de dicha acción. Usamos una función de recompensa relacionada a la \textit{fidelidad} de transmisión de información del sistema influenciado por las diferentes acciones, la cual nos guía en ese proceso. El estado obtenido es el vector de estado de la cadena de qubits. El aprendizaje se logra entrenando el agente usando \textit{diferencia temporal} (QN) en una versión \textit{profunda} (Deep QN) y con una política de exploración $\epsilon-greedy$. La \textit{función de valor} es aproximada por una red neuronal \textit{fully connected} que es entrenada por \textit{lotes} a partir de la experiencia generada en el proceso y que recibe como entrada el estado del entorno y como salida los valores aproximados de la función de valor para ese estado en cada una de las acciones (\cite{sutton_barto_2018}). 

DRL es cada vez más utilizado para estudiar el control de sistemas cuánticos, ver referencias (\cite{quantumrl}) y (\cite{quantumrl2}). Una descripción más detallada del algoritmo usado puede encontrarse en (\cite{Zhang2018}) y sus respectivas citas. La versión empleada en este trabajo sigue el patrón básico descrito en (\cite{implementacion_qlearning}). Para la implementación de las redes neuronales profundas necesarias se utilizó TensorFlow. Nuestra implementación necesita una serie de hiperparámetros que definen la arquitectura, conectividad y número de neuronas por capa, además de los parámetros necesarios en el Q-learning, el learning rate, $\alpha$, el discount factor. $\gamma$, y la función de recompensa (ver \href{https://github.com/sofips/JAIIO_material_suplementario/blob/main/drl_details.md}{aquí}).

\vspace{4pt}
\noindent{\bf Algoritmo genético}
\vspace{4pt}

Los Algoritmos Genéticos (AG) son métodos de optimización estocástica basados en poblaciones. Operan sobre una población $\mathcal{P}$ de individuos, donde cada individuo representa una solución candidata. La calidad de cada individuo se evalúa con una función de aptitud $f$. El propósito de la evolución es aumentar la aptitud (\cite{genetico}). 

Cada generación del AG consta de: (i) selección de individuos con probabilidades proporcionales a su aptitud; (ii) cruce de pares parentales mediante operadores de cruce, generando nuevos individuos; y (iii) mutación mediante un operador aleatorio que perturba genes seleccionados. La población se actualiza iterativamente, preservando los individuos más aptos, hasta que se cumple un criterio de convergencia (número máximo de generaciones o umbral de fidelidad).

Nuestro estudio considera como genes a las distintas acciones. Una secuencia completa de acciones representa un individuo de la población. A su vez, la aptitud de cada individuo, es decir, su \textit{fitness} está representada por una función proporcional a la fidelidad de transmisión. En particular, para comparar con el algoritmo de Deep Reinforcement Learning, diseñamos una \textit{fitness} inspirada en la función recompensa usada por \cite{Zhang2018} de este tipo de algoritmos que también es proporcional a la probabilidad de transición dada por \eqref{trans_prob}. 

Dado que hay un número discreto de acciones posibles, elegimos una mutación de tipo \textit{swap} que intercambia los valores de dos genes elegidos de forma aleatoria. Respecto al cruce, elegimos usar uno de tipo uniforme que intercambia valores aleatoriamente entre los individuos parentales en cada uno de los sitios. 

Los parámetros usados, además de la {\it fitness} utilizada pueden consultarse \href{https://github.com/sofips/JAIIO_material_suplementario/blob/main/ga_details.md}{aquí}.

\section{Resultados}

\begin{figure}
    \centering
    \includegraphics[width=0.43\textwidth]{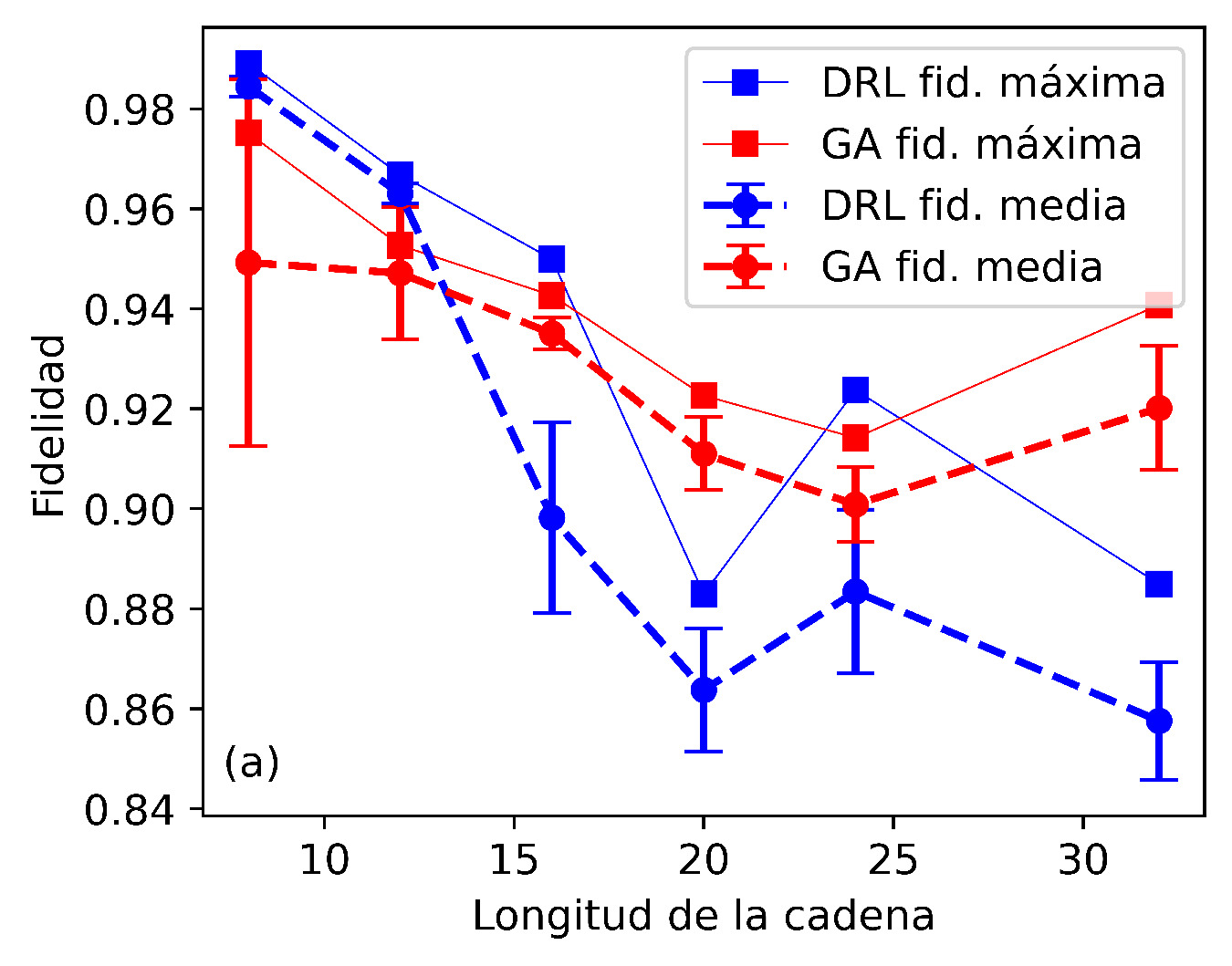}
      \includegraphics[width=0.43\textwidth]{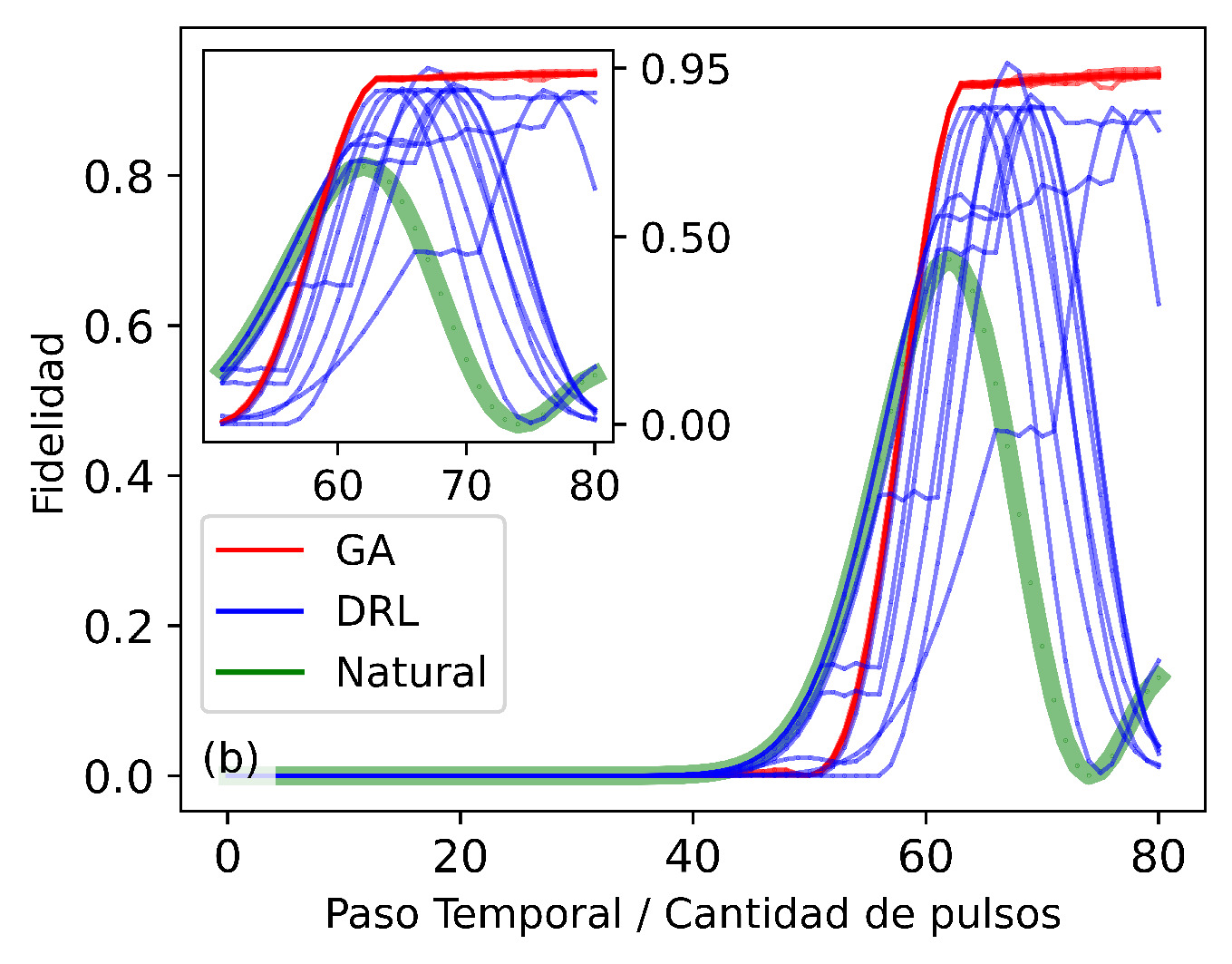}
    \caption{Panel a): Comparación de los valores de fidelidad obtenidos usando DRL (azul) y algoritmos genéticos (rojo). Teniendo en cuenta que ambos métodos son no deterministas, se toma un promedio sobre 10 ejecuciones. En el caso de DRL, se toman las soluciones correspondientes a los 10 valores más altos de fidelidad logrados por el agente.  En el caso del algoritmo genético, se ejecuta el algoritmo 10 veces obteniendo 10 soluciones distintas e independientes. Las curvas se corresponden con los valores de fidelidad medios sobre las 10 ejecuciones y con la fidelidad máxima obtenida en los 10 intentos. Se utiliza mutación de tipo swap, crossover uniforme y una población de 2048 individuos, de la cual se seleccionan 205 padres en cada generación usando \textit{'steady-state-selection'}. Para el algoritmo de DRL, se utilizan los híper-parámetros del trabajo \cite{Zhang2018}. Panel b): Se muestra la evolución temporal de la probabilidad de transición para una cadena de 16 espínes para la mejor solución obtenida con ambos algoritmos. Se comparan con la evolución natural (verde), es decir, sin forzamientos.  }
    \label{fig:mi_figura}
\end{figure}

Además de reproducir los resultados del trabajo de (\cite{Zhang2018}), gracias a que los autores proveyeron algunos de sus códigos fuente, exploramos distintos conjuntos de parámetros a fin de mejorar sus resultados. En dicho trabajo el DRL es usado para detectar el {\em quantum speed limit}, no para obtener la máxima transmisión de estados posible. Se pudo comprobar que los resultados para la fidelidad de transmisión no son del todo buenos y que el agente no queda entrenado para proveer secuencias que resulten en una buena transmisión. Sin embargo el presente estudio sugiere que cambiando acciones, hiper parámetros y recompensa puede ser posible mejorar los valores de la probabilidad de transmisión.  Para el AG usamos la librería PyGAD (\cite{PyGAD}). La figura \ref{fig:mi_figura} muestra la comparación de la probabilidad de transmisión obtenida usando ambos métodos en cadenas de qubits de distinto largo.

En el panel a) se observa que para longitudes pequeñas el DRL parece la mejor opción, aunque para longitudes largas el AG muestra claras ventajas. Estas aumentan cuando se observa el panel b). El DRL produce "picos" en la probabilidad, lo cual limita el tiempo durante el cual la información puede ser adquirida. mientras que el AG obtiene dichos valores durante intervalos considerables, como se muestra en el panel b). 

Estos resultados preliminares son interesantes y muestran la necesidad de explorar más la influencia de los hiperparámetros en los resultados obtenidos. Las máximas probabilidades alcanzadas por ambos algoritmos dependen fuertemente de estas elecciones. Además, utilizando distintas funciones \textit{fitness} (AG) o \textit{recompensas} (DRL) es posible controlar propiedades físicas de la solución obtenida, como su localización. Por otro lado, estamos analizando el efecto de considerar otros grupos de acciones, por ejemplo, utilizando solo controles individuales en cada sitio de la cadena. 

Cabe destacar que el algoritmo genético es considerablemente más eficiente en términos de costo computacional ya que, al estar basado en evaluaciones independientes de las distintas secuencias, fue posible paralelizarlo para ejecutar las simulaciones en GPU. De esta manera, pueden obtenerse secuencias que provean fidelidades superiores a un 95 \% en minutos. Actualmente, estamos trabajando en formas de optimizar también la implementación del algoritmo de DRL.

Otros gráficos, y algunas extensiones pueden encontrarse en \cite{repo-github}.

\bibliographystyle{unsrtdin}
\bibliography{references}

\begin{thebibliography}{11}


\providecommand{\url}[1]{\texttt{#1}}
\expandafter\ifx\csname urlstyle\endcsname\relax
  \providecommand{\doi}[1]{doi: #1}\else
  \providecommand{\doi}{doi: \begingroup \urlstyle{rm}\Url}\fi

\bibitem[1]{Peron2025}
\textsc{Perón~Santana}, S ; \textsc{Domínguez}, M  ; \textsc{Osenda}, O:
\newblock Quantum state transfer performance of Heisenberg spin chains with
  site-dependent interactions designed using a generic genetic algorithm.
\newblock {In: }\emph{Physica Scripta} 100 (2025), apr, Nr. 5, S. 055110

\bibitem[2]{Zhang2018}
\textsc{Zhang}, Xiao-Ming ; \textsc{Cui}, Zi-Wei ; \textsc{Wang}, Xin  ;
  \textsc{Yung}, Man-Hong:
\newblock Automatic spin-chain learning to explore the quantum speed limit.
\newblock {In: }\emph{Phys. Rev. A} 97 (2018), S. 052333

\bibitem[3]{drl}
\textsc{François-Lavet}, Vincent ; \textsc{Henderson}, Peter ; \textsc{Islam},
  Riashat ; \textsc{Bellemare}, Marc~G.  ; \textsc{Pineau}, Joelle:
\newblock \emph{An Introduction to Deep Reinforcement Learning}.
\newblock Now Publishers, 2018.
\newblock \url{http://dx.doi.org/10.1561/2200000071}.
\newblock \url{http://dx.doi.org/10.1561/2200000071}

\bibitem[4]{sutton_barto_2018}
\textsc{Sutton}, Richard~S. ; \textsc{Barto}, Andrew:
\newblock \emph{Reinforcement learning: an Introduction}.
\newblock Cambridge, Ma ; London : MIT Press, 2018. --
\newblock  33–39 S. --
\newblock ISBN 9780262039246

\bibitem[5]{quantumrl}
\textsc{Bukov}, Marin ; \textsc{Day}, Alexandre G.~R. ; \textsc{Sels}, Dries ;
  \textsc{Weinberg}, Phillip ; \textsc{Polkovnikov}, Anatoli  ; \textsc{Mehta},
  Pankaj:
\newblock Reinforcement Learning in Different Phases of Quantum Control.
\newblock {In: }\emph{Phys. Rev. X} 8 (2018), Sep, 031086.
\newblock \url{http://dx.doi.org/10.1103/PhysRevX.8.031086}. --
\newblock DOI 10.1103/PhysRevX.8.031086

\bibitem[6]{quantumrl2}
\textsc{Sgroi}, S ; \textsc{Zicari}, G ; \textsc{Imparato}, A  ;
  \textsc{Paternostro}, M:
\newblock A reinforcement learning approach to the design of quantum chains for
  optimal energy and state transfer.
\newblock {In: }\emph{Machine Learning: Science and Technology} 6 (2025), jan,
  Nr. 1, 015012.
\newblock \url{http://dx.doi.org/10.1088/2632-2153/ada71d}. --
\newblock DOI 10.1088/2632--2153/ada71d

\bibitem[7]{implementacion_qlearning}
\textsc{Doshi}, Ketan:
\newblock \emph{Reinforcement Learning Explained Visually (Part 5): Deep Q
  Networks}.
\newblock \url{https://tinyurl.com/336cjsux}.
\newblock \,Version:\,Feb 2021

\bibitem[8]{genetico}
\textsc{Mirjalili}, Seyedali:
\newblock \emph{Evolutionary Algorithms and Neural Networks : Theory and
  Applications}.
\newblock Cham : Springer International Publishing, 2019. --
\newblock ISBN 9783319930251

\bibitem[9]{PyGAD}
\textsc{Gad}, Ahmed~F.:
\newblock PyGAD: an intuitive genetic algorithm Python library.
\newblock {In: }\emph{Multimedia Tools and Applications} 83 (2024), Jun, Nr.
  20, S. 58029--58042. --
\newblock ISSN 1573--7721

\bibitem[10]{repo-github}
\textsc{Perón~Santana}, Sofía:
\newblock \emph{Material Suplementario JAIIO (github)}.
\newblock \url{https://github.com/sofips/JAIIO_material_suplementario}.
\newblock \,Version:\,2025

\end{thebibliography}

\end{document}